\documentclass[sigconf]{acmart}

\AtBeginDocument{%
  \providecommand\BibTeX{{%
    \normalfont B\kern-0.5em{\scshape i\kern-0.25em b}\kern-0.8em\TeX}}}

\copyrightyear{2023}
\acmYear{2023}
\setcopyright{rightsretained}
\acmConference[IUI '23 Companion]{28th International Conference on Intelligent User Interfaces}{March 27--31, 2023}{Sydney, NSW, Australia}
\acmBooktitle{28th International Conference on Intelligent User Interfaces (IUI '23 Companion), March 27--31, 2023, Sydney, NSW, Australia}
\acmDOI{10.1145/3581754.3584133}
\acmISBN{979-8-4007-0107-8/23/03}

\begin{document}

\title[Fatigue and mental underload further pronounced in L3 driving]{Fatigue and mental underload further pronounced in L3 conditionally automated driving: Results from an EEG experiment on a test track}

\author{Nikol Figalová}
\email{nikol.figalova@uni-ulm.de}
\orcid{0000-0001-7618-4852}
\affiliation{%
  \institution{Deptartment of Clinical and Health Psychology, Ulm University}
  \streetaddress{Albert-Einstein-Allee 45}
  \city{Ulm}
  \country{Germany}
  \postcode{89081}
}

\author{Hans-Joachim Bieg}
\orcid{0000-0002-3291-2683}
\affiliation{%
  \institution{Robert Bosch GmbH}
  \city{Renningen}
  \country{Germany}}

    \author{Michael Schulz}
    \orcid{0000-0001-6139-5382}
\affiliation{%
  \institution{Robert Bosch GmbH}
  \city{Renningen}
  \country{Germany}}
  
\author{Jürgen Pichen}
\orcid{0000-0003-2844-0465}
\affiliation{%
  \institution{Department of Human Factors, Ulm University}
  \city{Ulm}
  \country{Germany}}

\author{Martin Baumann}
\orcid{0000-0002-2668-2527}
\affiliation{%
  \institution{Department of Human Factors, Ulm University}
  \city{Ulm}
  \country{Germany}}

  \author{Lewis Chuang}
\orcid{0000-0002-1975-5716}
\affiliation{%
  \institution{Department of Human and Technology, Institute for Media Research, Faculty of Humanities, Chemnitz University of Technology}
  \city{Ulm}
  \country{Germany}}

\author{Olga Pollatos}
\orcid{0000-0002-0512-565X}
\affiliation{%
  \institution{Deptartment of Clinical and Health Psychology, Ulm University}
  \city{Ulm}
  \country{Germany}}

\renewcommand{\shortauthors}{Figalová et al.}

\begin{abstract}

Drivers' role changes with increasing automation from the primary driver to a system supervisor. This study investigates how supervising an SAE L2 and L3 automated vehicle (AV) affects drivers' mental workload and sleepiness compared to manual driving. Using an AV prototype on a test track, the oscillatory brain activity of 23 adult participants was recorded during L2, L3, and manual driving. Results showed decreased mental workload and increased sleepiness in L3 drives compared to L2 and manual drives, indicated by self-report scales and changes in the frontal alpha and theta power spectral density. These findings suggest that fatigue and mental underload are significant issues in L3 driving and should be considered when designing future AV interfaces.

\end{abstract}



\begin{teaserfigure}
  \includegraphics[width=\textwidth]{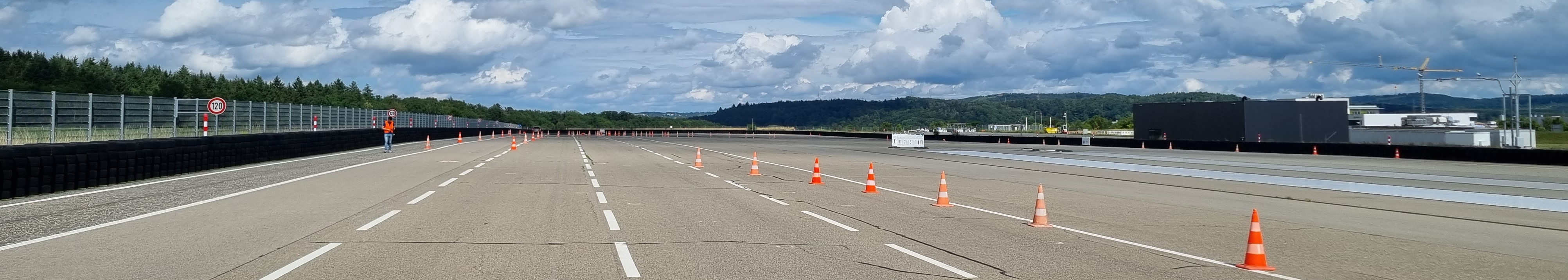}
  \caption{The test track which was used in the experiment.}
  \Description{Landing strip of an old airport which was re-built into a test-track.}
  \label{fig:teaser}
\end{teaserfigure}

\received{}
\received[]{}
\received[]{}

\maketitle
\section{Introduction}

With increasing vehicle automation, drivers' role changes. While L0 SAE \cite{SAEJ3016} vehicles are operated solely by the driver, SAE L2 vehicles can take over the driving task when certain conditions are met. Nevertheless, L2 vehicles must be constantly monitored. This is no longer necessary on SAE L3 - drivers may engage in non-driving-related activities while the vehicle operates in an automated mode. This will likely result in a significant shift in the skills and abilities drivers need to possess.

The success of the transition to automated vehicles (AVs) depends on drivers' ability to adapt to their new role as monitors rather than primary operators. However, drivers may not be well prepared for this change \cite{bengler2014three, merat2009drivers, brandenburg2014switching}. Automated driving systems can perform the active tasks of steering and navigating complex traffic situations, while drivers are expected to passively supervise the system. This can lead to mental underload \cite{stapel2019automated, mcwilliams2021underload}. The lack of stimulation makes highly automated driving difficult to satisfy psychological fulfilment and user experience \cite{frison2019you}, resulting in drivers being unable to stay alert during extended periods of automated driving \cite{vogelpohl2019asleep, bieg2020task}. Therefore, AV drivers experience more fatigue compared to those driving manually \cite{kundinger2020driver, michael2006sustained}.

Fatigue significantly impairs the driver's ability to operate a vehicle safely \cite{kumari2017survey, maclean2003hazards, higgins2017asleep, markkula2017simulating}, hinders their attention \cite{boksem2005effects}, and situation awareness \cite{bongo2022effect}. Moreover, drivers tend to misuse or abuse the AV system, either intentionally or due to a lack of understanding of their tasks as AV operators \cite{robinson_2023, parasuraman1997humans}. These issues should be addressed when designing an intelligent user interface (UI) that monitors and supports drivers' performance \cite{kohn2019improving, figalova2022ambient, wilbrink2020reflecting, sucu2013haptic} and manages their attention \cite{wintersberger2018let, wintersberger2019attentive, bailey2006need}. 

Drivers fatigue can be detected using computer vision \cite{xiao2019fatigue, dwivedi2014drowsy, vural2008automated} and artificial intelligence \cite{karwowski2020artificial, vural2007drowsy} using measures such as eyelid closure \cite{zhou2020driver, chang2014driver}, electrocardiogram \cite{jung2014driver}, or electroencephalography (EEG) \cite{morales2017monitoring, salimuddin2018driver, hu2018automated}. Most EEG studies report an increase in frontal alpha activity as an indicator of fatigue, while an increase in frontal theta power indicates a higher mental workload (for an overview, see Lohani et al. \cite{lohani2019review}). Detecting the driver's state is crucial for adaptive UIs. However, we should use a prophylactic approach to design safe, desirable, and acceptable intelligent UIs. Therefore, we must understand what drivers experience while monitoring the AV to design UIs that meet their needs.

Most research on driver fatigue and mental underload has been conducted using driving simulators \cite{de2014effects}, which have limitations that may affect the relevance of findings in the real world \cite{hock2018design}. Moreover, most studies used scenarios that differ from what drivers would encounter in actual AVs (e.g., critical situations are rather unusual in natural traffic conditions \cite{klischat2019generating}, although they occur relatively frequently in simulator experiments). We propose an experiment using a test track and an AV prototype to address these issues. Participants will experience a monotonous, non-stimulating task similar to operating real L2 and L3 AV. We will measure sleepiness and mental workload using self-report scales and EEG. The goal is to compare drivers' cognitive states when interacting with L2 and L3 automation in a realistic setting. Our results will serve as a knowledgebase for the designers of future intelligent UIs for highly automated vehicles.

\section{Method}

\subsection{Apparatus and experimental conditions}
We used an AV prototype (VW Golf 7) at a test track in Stuttgart, Germany (Figure \ref{fig:teaser}). The AV travelled in loops of 2000 m at a speed of 50 km/h on straight sections and 20 km/h in curves. The EEG was recorded using 32 channels placed according to the international 10-20 system. We used active electrodes and kept the impedance below $25 k\Omega$. The data were recorded with a 1000 Hz sampling rate. The perceived mental workload was assessed using the NASA-TLX \citep{hart2006nasa}. Sleepiness was measured using the Karolinska sleepiness scale \cite{shahid2011karolinska}. 

Participants experienced three counterbalanced conditions (17 minutes each): monitoring an SAE L3 vehicle; driving an SAE L2 vehicle; manual driving. Participants were primed about the differences in their tasks in each condition. In the \textbf{L3 ride}, participants were told that the vehicle would request them to take control in advance if necessary. In the \textbf{L2 ride}, participants were told to supervise the automated system constantly and intervene in case of failure. In the \textbf{manual ride}, participants were asked to drive at 50 km/h on straight portions and slow to 20 km/h in curves.

\subsection{Participants}
We recruited 23 participants (14 females; \textit{M} = 41.24 years; \textit{SD} = 14.71) with no prior AV experience. All participants had normal or corrected-to-normal vision, no known neurological or psychiatric disease, and a valid German driving license (on average for \textit{M} = 18.12 years, \textit{SD} = 12.63). All participants provided informed consent and received 90 Euros.

\subsection{EEG signal processing and analysis}
The EEG data were preprocessed in Matlab version R2022a according to the BeMoBil pipeline \citep{klug2022bemobil}. The cleaned data were bandpass-filtered between 0.5 and 30 Hz. The relative spectral power density was analyzed in the theta band (4-8 Hz) and the alpha band (8 to 12 Hz) on the Fz electrode.

\section{Results}
\subsection{Self-report mental workload and sleepiness}
A repeated measures ANOVA with a Greenhouse-Geisser correction determined that the mean NASA-TLX score differed between the rides (\textit{F}(1.569, 34.514) = 7.313, \textit{p}=.004, $\omega_{}^{2}$ = 0.070). Post-hoc testing using the Bonferroni correction revealed that the mental workload was lower in the L3 rides compared to the L2 rides (mean difference=10.957, \textit{p}=.020, \textit{d}=.537), and in the L3 rides compared to the manual rides (mean difference=14.043, \textit{p}=.002, \textit{d}=.688). 

\begin{figure}[h]
  \centering
  \includegraphics[width=1\linewidth]{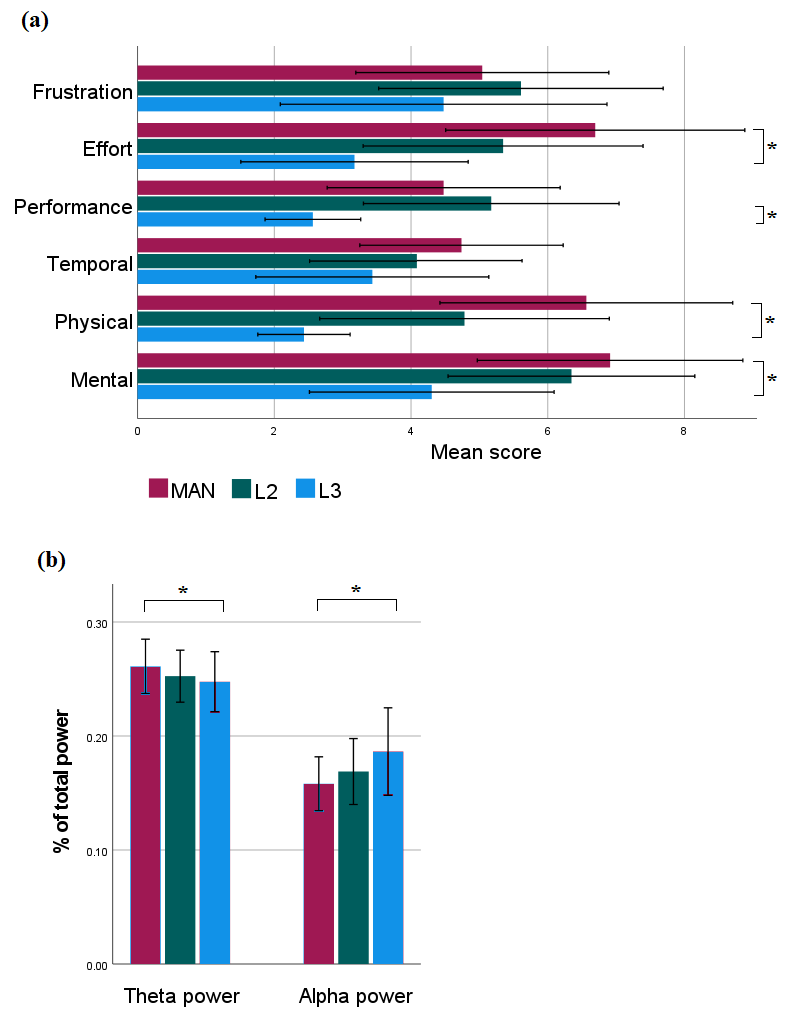}
  \caption{\textbf{(a)} The mean scores of the six dimensions of mental workload measured by NASA-TLX after each ride.\textbf{ (b)} The relative power spectral density measured on the Fz electrode in the theta (4-8 Hz) and alpha (8-12 Hz) band. Significant differences in the mean scores are marked with an asterisk. The error bars represent the 95\% confidence interval.}
  \Description{The mean scores of the six dimensions of NASA-TLX and the comparison of alpha and theta power in different conditions.}
  \label{graphs}
\end{figure}

Figure \ref{graphs}a presents the mean score of the six dimensions of NASA-TLX measured after the L2, L3, and manual rides. The scores suggest higher mental, physical, and effort demands in the manual rides compared to the L3 rides. Furthermore, we found a higher performance demand for the L2 rides compared to the L3 rides. 
Moreover, we compared the self-report sleepiness. A repeated measures ANOVA with a Greenhouse-Geisser correction determined that the mean KSS score differed between the rides (\textit{F}(1.961, 43.147) = 3.386, \textit{p}=.044, $\omega_{}^{2}$ = 0.019). Post-hoc testing using the Bonferroni correction revealed that drivers experienced more sleepiness in the L3 rides compared to the manual rides (mean difference=0.826, \textit{p}=.039, \textit{d}=.407). 

\subsection{EEG measures}
We assessed the power spectral density in the theta (4-8 Hz) and alpha (8-12 Hz) bands. Results are visualised in \ref{graphs}b. A repeated measures ANOVA determined that the theta power differed between the rides (\textit{F}(2.000, 42.000) = 4.937, \textit{p}=.012, $\omega_{}^{2}$ = 0.008). Post-hoc testing using the Bonferroni correction revealed that the theta power was higher in the manual rides compared to the L3 rides (mean difference=0.013, \textit{p}=.010, \textit{d}=.246), which suggests that drivers experienced higher mental workload when driving manually. 

Moreover, a repeated measures ANOVA determined that the alpha power differed  between the rides (\textit{F}(2.000, 42.000) = 4.417, \textit{p}=.018, $\omega_{}^{2}$ = 0.021). Post-hoc testing using the Bonferroni correction revealed that the alpha power was higher in the L3 rides compared to the manual rides (mean difference=0.028, \textit{p}=.016, \textit{d}=.405), which suggests that drivers experienced more fatigue and drowsiness during the L3 rides. 

\section{Discussion}
The present study reports preliminary findings on the perception of L3 driving among drivers in a real AV. Our results suggest that self-reported mental workload was lowest during L3 rides, with no difference in mental workload between L2 and manual rides. The decreased frontal theta brain activity during L3 rides supports the idea that drivers experience a low mental workload while passively monitoring L3 AVs. Moreover, we found no difference between manual and L2 rides but a significant difference between manual and L3 rides. The increased alpha activity during L3 rides supports the argument that drivers become more fatigued and drowsy while passively monitoring L3 AVs. 

Our findings are consistent with previous research indicating that vehicle automation reduces mental workload \cite{kundinger2020driver, de2014effects, mcwilliams2021underload, stapel2019automated, bieg2020task} and increases drowsiness \cite{kundinger2020driver, michael2006sustained, vogelpohl2019asleep}. However, previous studies primarily focused on L2 driving, and only scarce evidence has been published about L3 AVs outside the driving simulator. Our results suggest that the underload effect on mental workload may be further pronounced in L3 driving, which can be potentially dangerous given that L3 drivers are still partially responsible for the driving task \cite{SAEJ3016}. This study was conducted on a test track with automation inexperienced drivers. Real traffic environment, as well as the effect of long-term experience with automation, should be addressed in future studies. 

In conclusion, our experiment highlights the importance of designing UIs for L3 vehicles with consideration for drowsy and inattentive drivers. Dynamic UIs that manage drivers' attention and adapt to the current driving context are recommended. Issues such as attention shifting and cue saliency must be addressed. A driving monitoring system that detects inattention and engages with the driver in a timely manner ahead of a control transition should be included in the design of safe and desirable UIs for L3 vehicles. Further research is necessary to provide more specific recommendations addressing sleepiness and mental underload in the design of intelligent UIs for L3 AVs. 

\begin{acks}
  This project has received funding from the European Union's Horizon 2020 research and innovation programme under the Marie Sklodowska-Curie grant agreement 860410 and was supported by the German Federal Ministry for Economic Affairs and Climate Action (Bundesministerium für Wirtschaft und Klimaschutz, BMWK) through the project RUMBA (projekt-rumba.de). We would like to thank Philipp Alt and Erdi Kenar from Robert Bosch GmbH for their involvement in the project. 
\end{acks}

\bibliographystyle{ACM-Reference-Format}
\bibliography{Z_references}

\end{document}